\documentclass[letterpaper, 10 pt, conference]{ieeeconf} 
\IEEEoverridecommandlockouts                              
\usepackage{xcolor}
\usepackage{amsmath}
\usepackage{graphicx}
\usepackage{soul}
\usepackage{subcaption}
\usepackage{bm}
\usepackage{amsmath}
\usepackage{amssymb}
\usepackage{soul}

\newcommand{\real}{{\mathbb{R}}}

\usepackage{algorithm}
\usepackage{algorithmic}

\usepackage{graphicx}

\usepackage{amsmath}
\usepackage{amssymb}

\usepackage{verbatim}

\begin{document}
%

\title{\LARGE \bf Fuel and Time Optimal USV Trajectory Planning under Flexible Refueling Constraints}

%
%
%

\author{Aaron Kandel$^{1}$,
Chu Xu$^{2}$, 
David Cardona$^{3}$, and
        Carey Whitehair$^{4}$
\thanks{This work is supported by a National Science Foundation Graduate Research Fellowship.}
\thanks{$^{1}$Aaron Kandel is a PhD student at the Department of Mechanical Engineering at the University of California, Berkeley, Berkeley,
CA, 94704 USA  {\tt\small aaronkandel@berkeley.edu}.}%
\thanks{$^{2}$Chu Xu is a PhD Student with the Department of Mechanical Engineering at the University of Maryland, College Park, MD 20742 USA {\tt\small chuxu88@umd.edu}.}
\thanks{$^{3}$David Cardona is with the Department of Mechanical Engineering at The Pennsylvania State University, University Park, PA 16802 USA.}
\thanks{$^{4}$Carey Whitehair is a PhD student with the Department of Mechanical Engineering, University of Washington, Seattle, WA 98115 USA {\tt\small caw13@uw.edu}.}
}
\maketitle

\begin{abstract}

This paper addresses the problem of trajectory optimization for an unmanned surface vehicle while considering direction-dependent ocean currents and flexible refueling constraints. This work is motivated by the rising interest in developing autonomous navigation technology for commercial, scientific, and military applications for ocean-bound unmanned surface vehicles. Relevant literature on such vehicles has addressed energy-efficient trajectory optimization and time-efficient trajectory optimization. However, the application of trajectory optimization techniques which include refueling stops and multi-objective optimization remains relatively unexplored. We address this open challenge by formulating the trajectory design problem as a nonconvex mixed-integer optimization program with a multi-objective cost function.  Then, we apply dynamic programming to solve this optimization program for fuel and time optimal trajectories.  We synthesize these results into a series of Pareto fronts which demonstrates the tradeoff between fuel consumption and trip time for a prototypical route with direction-dependent ocean currents.  Furthermore, the optimal trajectories for which we solve illustrate the changes in the vehicle’s behavior as the fuel level becomes low, and as the vehicle encounters counterproductive ocean currents.  Our results indicate several meaningful insights about the overall trajectory optimization process.

\end{abstract}



%
\IEEEpeerreviewmaketitle

\section{Introduction}
This paper presents a Pareto analysis of time and fuel optimal trajectory planning for an autonomous ocean craft subject to direction-dependent ocean currents and flexible refueling constraints.

The study of unmanned surface vehicles (USV), specifically watercraft, spans several decades in the literature.  However, it was not until recently that USVs have become a viable option for diverse commercial, scientific, and military applications.  Relevant advances in controls and communications research are largely responsible for this development, enabling greater capabilities for autonomy and robust design.  Robust design of USVs is critical given the dynamic and vast ocean environments within which they operate.  For instance, in conventional military applications USVs are tasked with minesweeping missions, anti-submarine warfare, and surveying hazardous environments \cite{Liu00}.

Modeling the dynamics of USVs is a challenging task because of the complex set of forces acting on the vehicle, and the significant geometric dependencies of these forces.  Model complexity varies drastically, with the most complex and comprehensive dynamical models giving direct consideration to ocean swells and motion in 6 degrees of freedom \cite{Heins00}.  Other simple USV kinematics models utilize basic relations for surge, sway, and yaw movements \cite{Wang00}.  However as always, the best modeling approach is subject to the demands and circumstances placed upon the vehicle.  There has been significant work to design path planning for vehicles with slow dynamics such as underwater gliders and other vehicles that predominantly rely on ocean current for motion \cite{Yazdani00, Kim00}.  For these applications, it is critical that ocean current plays a strong role in the governing dynamics model of the vehicle.  For relevant estimation problems, ocean currents can also strongly impact both the quality of sensor measurements as well as potentially unmodeled disturbances \cite{Horner00}.

Past work has applied several tools from optimal control to address trajectory optimization problems which are immediately relevant to USVs. These optimization problems can take place over long-distance planning, or in over short distances to control events including docking maneuvers.  Due to the inherent non-linear nature of vehicle trajectory optimization, nonlinear programming (NLP) methods are frequently applied to solve these problems. One primary approach for time or energy efficient trajectory optimization has been implementing an initial trajectory based on global information and then updating this trajectory in real-time \cite{Jones00}. Work by Mills et al. has used concepts from differential flatness and pseudospectral methods to solve an energy-aware trajectory optimization problem for unmanned aerial vehicle as an NLP \cite{Mills00}. Similar research on unmanned underwater vehicles (UUVs) has explored the use of model predictive control (MPC). As demonstrated by Guerreiro et al., MPC is advantageous in the presence of disturbances such as ocean currents due to continuously updating the optimal path at each time step \cite{Guerreiro00}.  MPC is, however, often employed on-line at substantially increased computational time and effort. To overcome the computational cost of methods such as MPC, some have employed quasi-optimal approaches \cite{Yazdani00}. Obstacle avoidance also presents a unique challenge to trajectory optimization for USVs \cite{Wang00, Svec00}. Relevant methods which address obstacle avoidance problems frequently utilize tracking control to avoid collisions during the routing process.  

The presence of ocean currents as direction dependent disturbances poses a significant effect on the optimal trajectory a USV may take.  Work by Hirsch et al.  has elegantly addressed this challenge \cite{Dolinskaya00}. Ocean currents also can have a significant effect on the energy expenditure of a USV.  As a result, fuel-optimal strategies for USV trajectory optimization have been observed to exploit ocean currents to a similar degree.  Other research by Kim et al. has applied genetic optimization to explore the impact of environmental loads on the optimal trajectory taken by a USV \cite{Kim01}.

While time and fuel optimal USV trajectory optimization has been studied throughout the literature, an analysis of the tradeoff between the two objectives is relatively unexplored.  This is especially true in the case where refueling constraints must be considered.  This work seeks to address these open challenges by analyzing USV trajectory optimization with respect to fuel consumption and trip time when refueling constraints must be considered. First, we formulate a discrete-time optimal control problem conducive for a conventional dynamic programming solver.  Through Pareto optimization techniques, we explore the competing fuel and trip time tradeoff in detail.  Furthermore, we vary the initial fuel tank level of the USV to explore its impact on the refueling behavior of the vessel.  We also examine the effect of direction dependent ocean currents on the fuel consumption and overall routing of the USV.

The remainder of this paper is organized as follows.  Section II describes our formal problem statement, including the model we use to approximate the USV dynamics.  Section III details our reformulation of the problem in discrete time, which allows us to apply a dynamic programming solution approach.   Section IV illustrates our optimal trajectory results, including our Pareto analysis of the fuel and trip time tradeoff. Finally, our conclusion contextualizes our findings and their implications.

\section{Problem Formulation}

This sections describes our formulation of the USV trajectory optimization problem.  In the first subsection, we detail the dynamical model of the USV.  Then, we provide an explicit formulation of the optimal control problem statement.  Finally, we describe the ocean environment within which we optimize the vessel's trajectories subject to varying refueling requirements.

\subsection{USV Dynamical Model}
To solve the time and fuel optimal USV trajectory planning problem, we employ a relatively simple dynamical model of an ocean craft.  Namely, we assume that the vehicle is operating at steady-state, or that it operated on a timescale long enough such that we can neglect transient dynamics from propulsion and drag forces.  We do this so our representation of the dynamics can be addressed with a dynamic programming-based solver which possesses limitations regarding time and space complexity.  Our nonlinear 3rd-order dynamical model takes the following form:
\begin{multline} \label{eqn::dyn1}
    \dot{x}(t) = \begin{pmatrix}
 u_1(t)\\
u_2(t)  \\
u_3(t) - \dot{m}_{f}(t)
\end{pmatrix} + \begin{pmatrix}
1\\1\\
0\\
\end{pmatrix}*v_{cur}(x(t))
\end{multline}
where $x(t) \in \real^3$ is the state vector at time $t$, $x_1(t)$ and $x_2(t)$ are the east-west and north-south positions of the vehicle, $x_3(t)$ is the current level of fuel in the vehicle's fuel tank, $u_1(t)$ and $u_2(t)$ are velocity inputs in the east-west and north-south directions, $u_3(t)$ is the input to the fuel tank level from refueling, $\dot{m}_{fuel}(t)$ is the rate of fuel consumption, and $v_{cur}(x(t))$ is the effect of the ocean current on the trajectory of the craft, where ocean current direction and magnitude are functions of the current position $x(t)$. The refueling input takes the form:
\begin{equation}\label{eqn::ncvx_con}
    u_3(t) = \left\{ \begin{array}{rcl}
u_{3;max} & \mbox{for} & {x},u \in \mathcal{P}\\ 
0 & \mbox{for} & {x},u \not\in \mathcal{P}\\
\end{array}\right.
\end{equation}
where $\mathcal{P}$ is the set of states and inputs such that the vehicle is stationary at one of the discrete port locations.  We assume that rate of fuel consumption is directly proportional to power. Consider the simple relation $P = Fv$, where $P$ is power, $F$ is force, and $v$ is velocity.  If we assume drag forces take the form $F_{drag} = \frac{1}{2} \rho C_d A_f v^2$ and that $v = \sqrt{(u_1(t))^2 + (u_2(t))^2}$, then $\dot{m} \propto{P}$ becomes:
\begin{equation}
    \begin{split}
    \dot{m}_f &= \alpha F_{drag}v \\
    &= \alpha \frac{1}{2} \rho C_d A_f v^3 \\
    &=\frac{1}{2} \alpha \rho C_d A_f \big{(}\sqrt{(u_1(t))^2 + (u_2(t))^2}\big{)}^3 \\
    &=\beta \big{(}(u_1(t))^2 + (u_2(t))^2\big{)}^{\frac{3}{2}} 
\end{split}
\end{equation}
We assume the ocean current does not affect the fuel consumption.  We base this assumption on the reasoning that the input velocity in our expression for $\dot{m}_f$ is applied relative to the ocean current.

\subsection{Optimal Control Problem Statement}
Our optimal control problem statement is as follows:
\begin{equation}\label{eqn::ct1}
    \underset{u \in \real^3}\min \int_0^T\big{[}(1-\lambda)\big{(}\frac{\dot{m}_{f}(t)}{\dot{m}_{fuel;max}}\big{)}+\lambda t_{inc} \big{]} dt 
\end{equation}
\text{Subject to:}
\begin{align}
\dot{x}(t) &= \begin{pmatrix}
 u_1(t)\\
u_2(t)  \\
u_3(t) - \dot{m}_{f}(t)
\end{pmatrix} + \begin{pmatrix}
1\\1\\
0\\
\end{pmatrix}*v_{cur}(x(t))\\
\dot{m}_{f}(t) &= \beta \big{(}(u_1(t))^2 + (u_2(t))^2\big{)}^{\frac{3}{2}} \\
x_{min} &\leq x(t) \leq x_{max}\\
u_{min} &\leq u(t) \leq u_{max}\\
u_3(t) &= \left\{ \begin{array}{rcl}
u_{3;max} & \mbox{for} & (x,u) \in \mathcal{P}\\ 
0 & \mbox{for} & (x,u) \not\in \mathcal{P}\\
\end{array}\right.\\
x(T) &\in \mathcal{X}_{terminal}
\end{align}

where 
\begin{equation}\label{eqn::ctend}
    t_{inc} = \left\{ \begin{array}{rcl}
1 & \mbox{for} & {x} \notin \mathcal{X}_{terminal}\\ 
0 & \mbox{for} & {x} \in \mathcal{X}_{terminal}\\
\end{array}\right.,
\end{equation}
$T$ is the final time, $\lambda$ is the scalarization weight which controls the tradeoff between fuel consumption and trip time, and $\mathcal{X}_{terminal}$ is the set of acceptable terminal states which only imposes conditions on the final x-y position of the vehicle.  We normalize the fuel consumption term in the cost function based on the fuel flow rate achieved when the velocity inputs are maximized.  An important constraint is that $x_{3;min} = 0$, meaning that the vehicle cannot run out of fuel at any point in its trip unless that point is at a refueling station or the final destination.  In Section III, we describe our discretization procedure which allows us to apply a standard dynamic programming algorithm to solve this optimal control problem.

\subsection{Ocean Environment}
We define the orientation of the time-invariant ocean currents based on arbitrary direction-dependent functions of both $x_1$ and $x_2$.  These nonconvex functions are given below:

\begin{equation}
    v_{cur;e}(x_1,x_2) = \gamma \sin\bigg(\frac{2\pi}{100}x_1\bigg)-\gamma \sin\bigg(\frac{2\pi}{100}x_2\bigg)
\end{equation}
\begin{equation}
    v_{cur;n}(x_2) = \frac{\gamma}{80}x_2 + \frac{1}{9}
\end{equation}
The vector $v_{cur}(x(t)$ now takes the form:
\begin{equation}
    v_{cur}(x(t)) = \begin{pmatrix}
 v_{cur;e}(x_1(t),x_2(t))\\
v_{cur;n}(x_2(t))  \\
0
\end{pmatrix}
\end{equation}
In this case we specifically limit the mean current velocity to approximately $5 km/hr$ with a maximum current velocity of approximately $9 km/hr$, in order to match observed averages among typical ocean currents \cite{NOAA00}. The overall state-space map of our problem, including the initial position and desired destination, is shown in Figure 1.  The vector field in Figure 1 describes the directions and relative magnitudes of the ocean currents at each position, and the refueling locations are marked with black squares.  We design this environment to create scenarios where the USV can exploit the ocean current speed and direction to either save fuel, or to reduce the overall travel time.

Table 1 provides a description of each parameter in problem.  We tune the fuel consumption proportionality parameter $\alpha$ such that our observations of fuel economy align with typical industry standards.
\begin{table}[h]
\caption{Model Parameters}
\label{table_example}
\begin{center}
\begin{tabular}{|c||c|c|}
\hline
Parameter & Description & Value\\
\hline
$x_{1;min}$ & Minimum $x_1$ & 0 $[km]$\\
\hline
$x_{1;max}$ & Maximum $x_1$ & 150 $[km]$\\
\hline
$x_{2;min}$ & Minimum $x_2$ & 0 $[km]$\\
\hline
$x_{2;max}$ & Maximum $x_2$ & 150 $[km]$\\
\hline
$x_{3;min}$ & Minimum $x_3$ & 0 $[Gal]$\\
\hline
$x_{3;max}$ & Maximum $x_3$ & 8 $[Gal]$\\
\hline
$u_{1;min}$ & Minimum $u_1$ & -21.33 $[\frac{km}{hr}]$\\
\hline
$u_{1;max}$ & Maximum $u_1$ & 21.33 $[\frac{km}{hr}]$\\
\hline
$u_{2;min}$ & Minimum $u_2$ & -21.33 $[\frac{km}{hr}]$\\
\hline
$u_{2;max}$ & Maximum $u_2$ & 21.33 $[\frac{km}{hr}]$\\
\hline
$u_{3;min}$ & Minimum $u_3$ & 0 $[\frac{Gal}{min}]$\\
\hline
$u_{3;max}$ & Maximum $u_3$ & 0.533 $[\frac{Gal}{min}]$\\
\hline
$C_d$ & USV Drag Coefficient & 0.4 $[-]$\\
\hline
$A_f$ & USV Frontal Area & 6 $[m^2]$\\
\hline
$\alpha$ & Power to Fuel Proportion & $1.2*10^{-7} [\frac{s^2}{m^2}]$\\
\hline
$\gamma$ & Ocean Current Coefficient & 0.0556 [-]
\end{tabular}
\end{center}
\end{table}

\begin{figure}
      \centering  
      \includegraphics[scale=0.57]{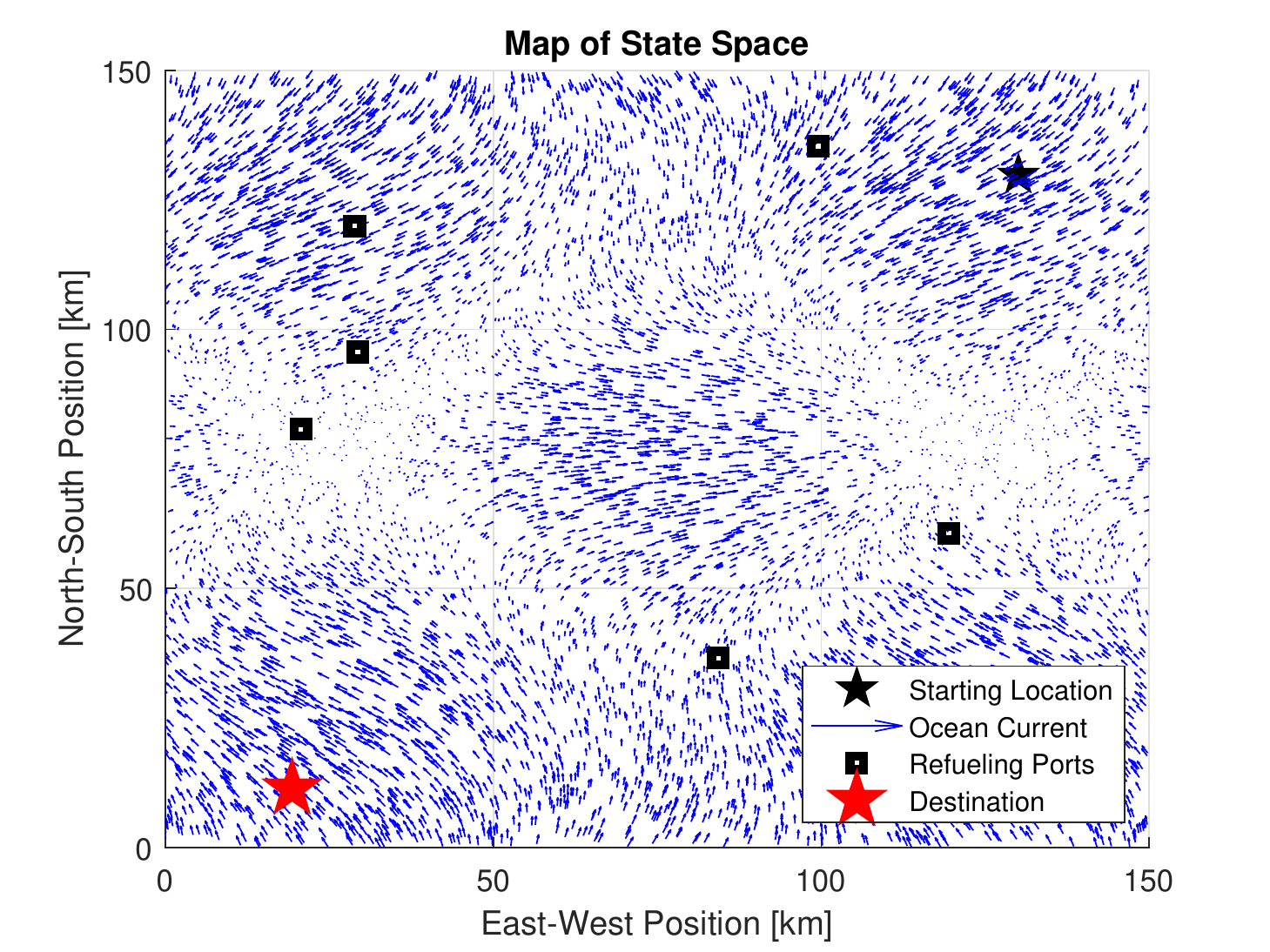}
      \caption{Map of Ocean Space}
      \label{figurelabel}
\end{figure}

\section{Dynamic Programming Solver}
The nonlinearity of the dynamical system model as well as the presence of the refueling constraint (\ref{eqn::ncvx_con}) make this a highly nonconvex mixed-integer program. Since our problem has a relatively low and tractable number of state and input variables, we elect to utilize a conventional dynamic programming approach to solve this problem across a finite time horizon.  To utilize a standard dynamic programming algorithm, we first discretize the continuous-time optimal control problem given by (\ref{eqn::ct1}-\ref{eqn::ctend}). Using a forward finite-difference method, the nonlinear system dynamics can be transformed into:
\begin{multline} \label{eqn::dt1}
    x(k+1) =  x(k) + \begin{pmatrix}
 u_1(k) \Delta t\\
u_2(k) \Delta t \\
u_3(k) - \dot{m}_{fuel}(k)
\end{pmatrix} \\+ \begin{pmatrix}
\Delta t\\
\Delta  t\\
0\\
\end{pmatrix}*v_{cur}(x(k))
\end{multline}
where $k$ is the current time and $\Delta t$ is the timestep length. The term $\dot{m}_{fuel}(k) = \dot{m}_f(t)* \Delta t$. For a large enough timestep, we now assume the control input $u_3$ is sufficient to refuel the USV's fuel tank in a single timestep.

Using a dynamic programming based solution approach requires we divide the state space and input space into a series of uniform discretized meshes.  Table 2 provides the parameters which define this discretization.  These align with our maximum and minimum values for $x$ and $u$ listed in Table 1.

Now, we can pose a discrete-time approximation of the continuous-time optimal control problem statement (\ref{eqn::ct1}-\ref{eqn::ctend}) as follows:
\begin{equation}
    \underset{u \in \real^3}\min \sum_{k=1}^N\big{[}(1-\lambda)\big{(}\frac{\dot{m}_{fuel}(k)}{\dot{m}_{fuel;max}}\big{)}+\lambda t_{inc} \big{]}  
\end{equation}
\text{Subject to:}
\begin{align}
&(\ref{eqn::dt1})\\
\dot{m}_{fuel}(k) &= \beta  \big{(}(u_1(k))^2 + (u_2(k))^2\big{)}^{\frac{3}{2}} \Delta t \\
x_{min} &\leq x(k) \leq x_{max}\\
u_{min} &\leq u(k) \leq u_{max}\\
u_3(k) &= \left\{ \begin{array}{rcl}
u_{3;max}\Delta t & \mbox{for} & (x,u) \in \mathcal{P}\\ 
0 & \mbox{for} & (x,u) \not\in \mathcal{P}\\
\end{array}\right.\\
x(N) &\in \mathcal{X}_{terminal}
\end{align}
where $N = \frac{T}{\Delta t}$ is the final timestep.  

\begin{table}[h]
\caption{Discretization Parameters}
\label{table_example}
\begin{center}  
\begin{tabular}{|c||c|}
\hline
Parameter & Value\\
\hline
$\Delta x_3$  &  0.1 $[Gal]$\\
\hline
$\Delta u_1$ & 10.67 $[\frac{km}{hr}]$\\
\hline
$\Delta u_2$ & 10.67 $[\frac{km}{hr}]$\\
\hline
$\Delta u_3$ & 8 $[\frac{Gal}{\Delta t}]$\\
\hline
$\Delta t$ & 15 $[min]$\\
\hline
$N$ & 65 $[Steps]$\\
\hline
\end{tabular}
\end{center}
\end{table}
To improve overall precision, we increase the relative coverage of our mesh by utilizing a Latin-hypercube sample to form the mesh of $x_1$ and $x_2$ points. Specifically, we generate a latin-hypercube sample ($n=7500$) within the bounds given by $x_{1;min}$, $x_{1;max}$, $x_{2;min}$, and $x_{2;max}$.  This overall discretization scheme necessitates implementing an interpolation rule when running the dynamic programming solver.  In this paper, we utilize a simple nearest-neighbor interpolation procedure.

\section{Results}
This section details our simulation results from our optimal trajectory planning.  The objective of this analysis is to explore the overall behavior of the USV when we vary several operating conditions of the optimal control problem.  These conditions include the initial fuel tank level and the scalarization parameter $\lambda$ which governs the tradeoff between fuel economy and trip time.

\subsection{Optimal Path Planning}
In this subsection, we combine our discretized problem structure with a general dynamic programming algorithm to solve for fuel and time optimal USV trajectories. Figures 3 and 4 demonstrate a series of optimal trajectories as we vary the initial fuel tank level and the scalarization parameter $\lambda$. 

\begin{figure}
      \centering  
      \includegraphics[scale=0.55]{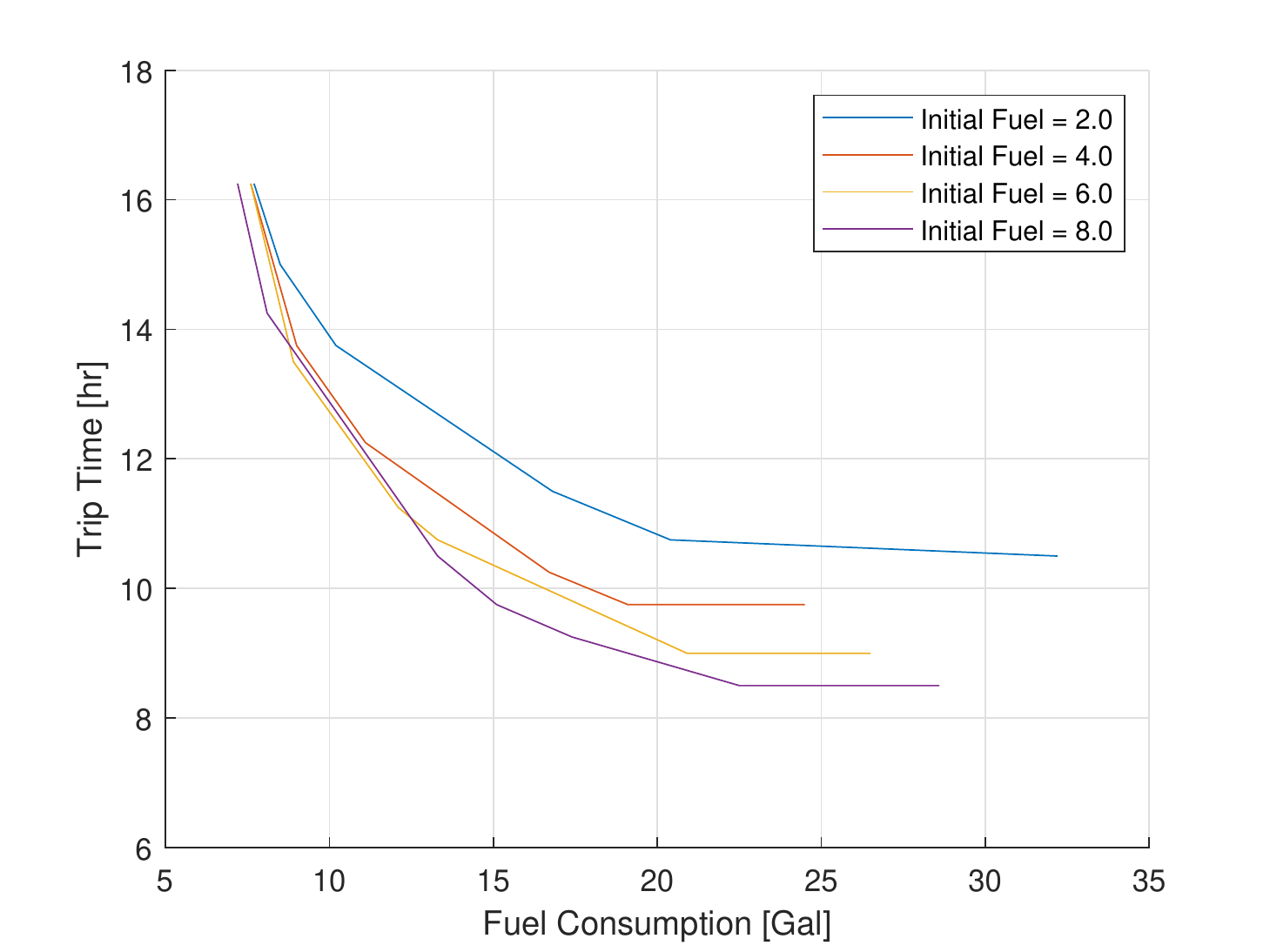}
      \caption{Pareto Fronts}
      \label{figurelabel}
\end{figure}
\begin{figure*}
      \centering  
      \includegraphics[trim = 10mm 5mm 15mm 5mm, clip, width=\textwidth]{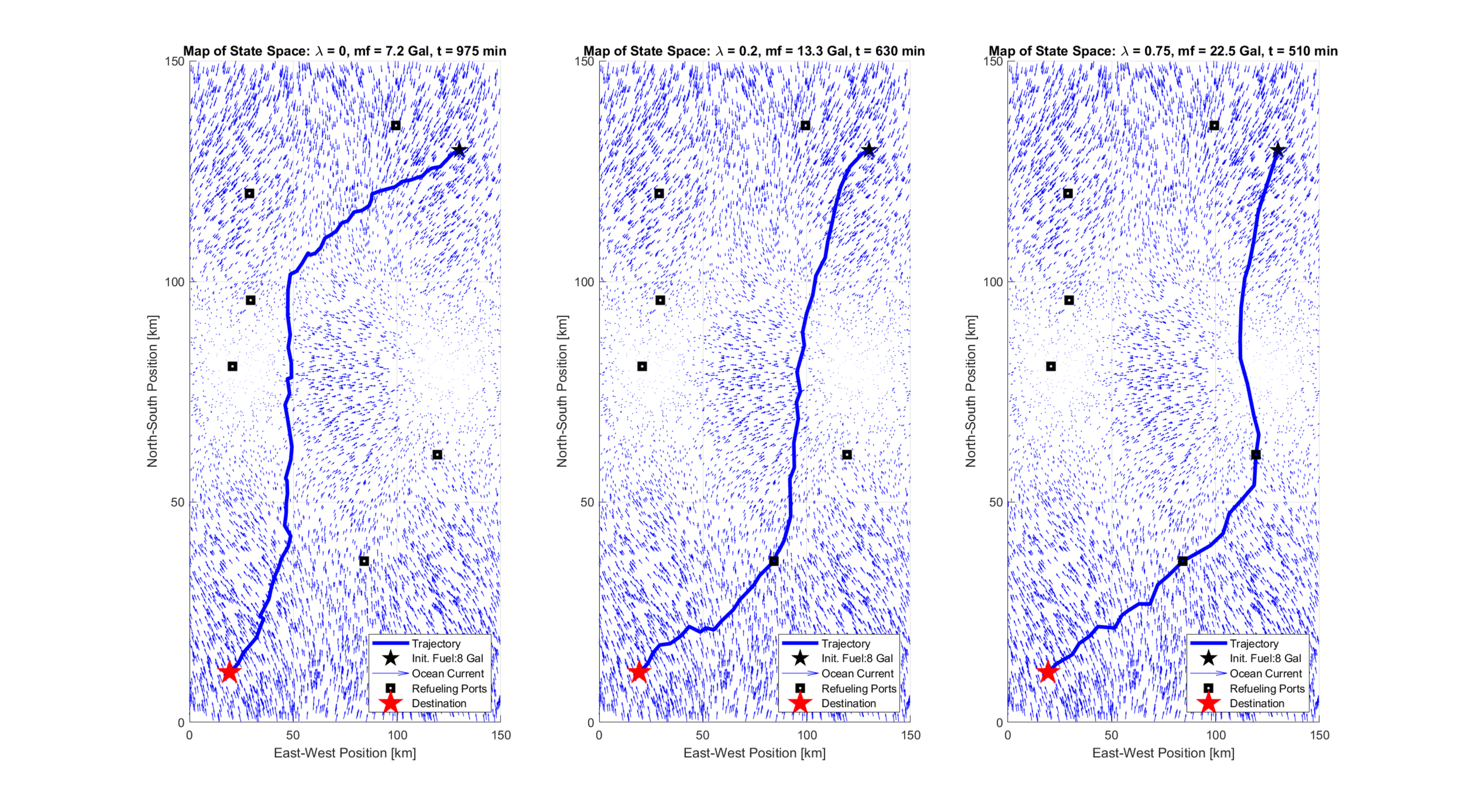}
      \caption{Comparison of optimal trajectories for initial fuel level of 8 gallons}
      \label{figurelabel}
\end{figure*}
\begin{figure*}
      \centering  
      \includegraphics[trim = 10mm 5mm 15mm 5mm, clip, width=\textwidth]{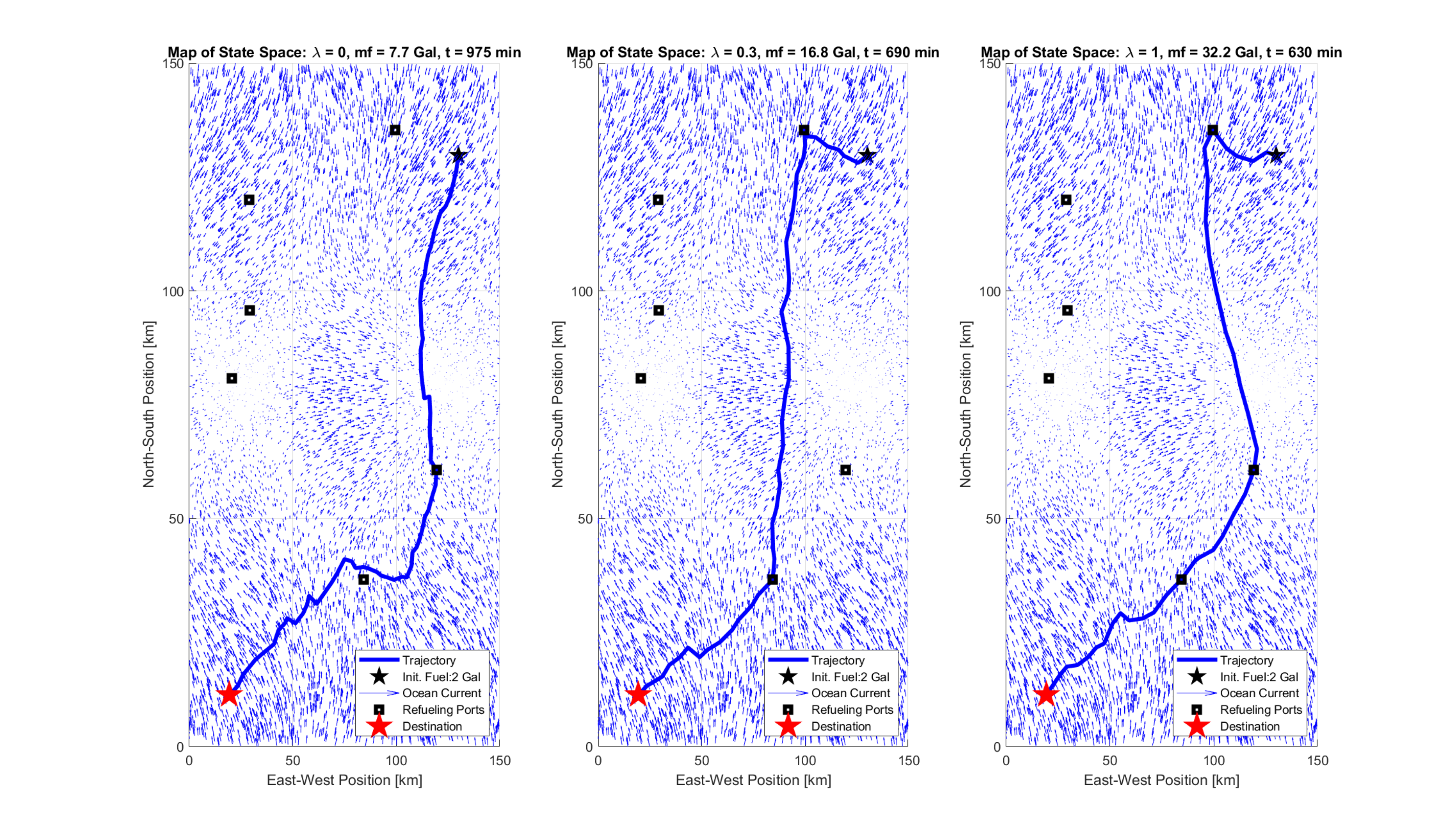}
      \caption{Comparison of optimal trajectories for initial fuel level of 2 gallons}
      \label{figurelabel}
\end{figure*}

Several insights are clearly visible from these results.  First, in the case where $\lambda=0$, the ocean current has much more visible effect on the trajectory the vehicle follows. The vehicle essentially allows the ocean current to do as much work as possible when fuel economy is the only consideration to the stage cost function. This is because the vehicle is defaulting to applying little to no input force whenever possible in order to save fuel.  This behavior can be seen in the leftmost portion of Figure 2, where the trajectory taken by the USV follows the ocean current vector field throughout a sizeable portion of its duration.  As $\lambda$ increases, this drifting becomes less and less noticeable.  Specifically, as the value of $\lambda$ increases the vehicle begins taking a more direct path to the desired destination.  This of course comes at the cost of additional fuel consumption, as the more direct trajectory necessitates neglecting and in some cases fighting directly against the direction of the ocean current.  As the value of $\lambda \rightarrow 1$, the vehicle will begin moving as fast as possible towards the destination as expected.  What is interesting is that this behavior requires the vehicle to in some cases refuel multiple times along its path.  In such cases, the vehicle defaults to roughly evenly spaced refueling ports which require the least detour from the most direct path.  


\subsection{Pareto Analysis}
This subsection details our comparative analysis between trip time and fuel consumption.  By varying the scalarization parameter $\lambda$, we can adjust the degree to which our optimal controller weighs fuel economy compared to overall trip time.  The objective of this analysis is to explore the relationship between these two aspects of the USV trajectory, and to see if certain rules and trends can be exploited to yield more efficient behavior.  

For our Pareto analysis, we vary the weighting parameter in the objective function $\lambda$ between zero and one in increments of 0.05.  Figure 2 shows the Pareto fronts obtained for four different initial fuel levels. We frequently find that some different values of $\lambda$ will share the same resulting trajectory.

This series of Pareto fronts yields several meaningful insights.  First, it is almost always true that having a higher initial fuel level yields significant improvement to both overall trip time and overall fuel consumption simultaneously.  This could be due to several factors.  In the event the vehicle needs to refuel, it will likely have to travel outside the path leading directly towards the desired destination.  This is, of course, dependent on the specific locations of the refueling ports.  The refueling itself also takes time, momentarily stopping the vehicle from making progress. Another key insight we can take from these Pareto fronts is that regardless of the initial fuel level, optimizing with the bare minimum of consideration to fuel economy yields essentially the same overall trip time but at significantly reduced fuel consumption.  In each of the initial fuel level cases, changing $\lambda$ from 1.0 to 0.95 had marginal change to trip time but decreased overall fuel consumption by 57.8\%, 28.3\%, 26.8\%, and 21.1\% (in increasing order of initial fuel level).

\section{Conclusion}

This paper presents a dynamic programming based analysis of multiobjective USV trajectory optimization.  Specifically, we explore the tradeoff between fuel consumption and overall trip time using Pareto optimization.  Our comparative analysis also examines how these competing objectives affect the trajectory taken by the vessel.  The principal novelty of our work comes from our implementation of refueling constraints, wherein the USV likely will need to stop to refuel on its way to its destination.  We augment this analysis with effects from direction-dependent ocean currents.

Our paper yields several meaningful insights. First, we find it is almost universally better in terms of fuel consumption and trip time for the USV to start its trip with as much fuel in its tank as possible.  We also find that it is typically better to give some, even minimal, consideration to fuel consumption compared to only considering overall trip time.  When minimal consideration is given to fuel consumption, we find the USV takes trajectories with nearly identical trip times to the time-greedy case but at between 20-50\% reduced fuel consumption.  Thus, this feature of a control algorithm can significantly extend the effective range of the USV. Furthermore, the relative effect of the ocean current on the shape of the optimal trajectory is considerable.  In general, the USV will avoid areas of high ocean current when the cost function heavily weights overall fuel economy.  This is true unless the direction of the ocean current aligns with the intended direction of the vessel.  In the same case, we frequently observe the USV being drawn seemingly off course by ocean currents.  However, the judicious use of control input allows the vessel to reach the destination with minimal overall fuel consumption.  

Another important note is that, in the event we wish to place the refueling ports in an optimal fashion, we expect higher average ocean currents will require a higher density of refueling ports. 


\bibliographystyle{./IEEEtran} 
\bibliography{./IEEEabrv,./bare_jrnl}

\begin{thebibliography}{10}
\providecommand{\url}[1]{#1}
\csname url@rmstyle\endcsname
\providecommand{\newblock}{\relax}
\providecommand{\bibinfo}[2]{#2}
\providecommand\BIBentrySTDinterwordspacing{\spaceskip=0pt\relax}
\providecommand\BIBentryALTinterwordstretchfactor{4}
\providecommand\BIBentryALTinterwordspacing{\spaceskip=\fontdimen2\font plus
\BIBentryALTinterwordstretchfactor\fontdimen3\font minus
  \fontdimen4\font\relax}
\providecommand\BIBforeignlanguage[2]{{%
\expandafter\ifx\csname l@#1\endcsname\relax
\typeout{** WARNING: IEEEtran.bst: No hyphenation pattern has been}%
\typeout{** loaded for the language `#1'. Using the pattern for}%
\typeout{** the default language instead.}%
\else
\language=\csname l@#1\endcsname
\fi
#2}}

\bibitem{Liu00}
Z.~Liu, Y.~Zhang, X.~Yu, and C.~Yuan, ``Unmanned surface vehicles: An overview
  of developments and challenges,'' \emph{Annual Reviews in Control}, vol.~41,
  pp. 71--93, 2016.

\bibitem{Heins00}
P.~H. Heins, B.~Ll.Jones, and D.~J. Taunton, ``Design and validation of an
  unmanned surface vehicle simulation model,'' \emph{Applied Mathematical
  Modeling}, vol.~48, pp. 749--774, 2017.

\bibitem{Wang00}
Y.~Wang, J.~Shen, and X.~Liu, ``Dynamic obstacles trajectory prediction and
  collision avoidance of usv,'' in \emph{Proceedings of the 2017 Chinese
  Control Conference}.\hskip 1em plus 0.5em minus 0.4em\relax Dalian, China:
  TCCT, 2017.

\bibitem{Yazdani00}
A.~Yazdani, K.~Sammut, A.~Lammas, and Y.~Tang, ``Real-time quasi-optimal
  trajectory planning for autonomous underwater docking,'' in \emph{Proceedings
  of the 2015 IEEE International Symposium on Robotics and Intelligent
  Sensors}.\hskip 1em plus 0.5em minus 0.4em\relax Langkawi, Malaysia: IEEE,
  2015.

\bibitem{Kim00}
K.~Kim and T.~Ura, ``Fuel-optimally guided navigation and tracking control of
  auv under current interaction,'' in \emph{Oceans 2003. Celebrating the Past
  ... Teaming Toward the Future}.\hskip 1em plus 0.5em minus 0.4em\relax San
  Diego, CA, USA: IEEE, 2003.

\bibitem{Horner00}
D.~Horner and M.~K. Mqana, ``Moving horizon estimation for undersea docking,''
  in \emph{Oceans 2017}.\hskip 1em plus 0.5em minus 0.4em\relax Aberdeen, UK:
  IEEE, 2017.

\bibitem{Jones00}
D.~Jones and G.~A. Hollinger, ``Planning energy-efficient trajectories in
  strong disturbances,'' \emph{IEEE Robotics and Automation Letters}, vol.~2,
  no.~4, pp. 2080--2087, 2017.

\bibitem{Mills00}
A.~B. Mills, D.~Kim, and E.~W. Frew, ``Energy-aware aircraft trajectory
  generation using pseudospectral methods with differential flatness,'' in
  \emph{Proceedings of the 2017 IEEE Conference on Control Technology and
  Applications (CCTA)}.\hskip 1em plus 0.5em minus 0.4em\relax Mauna Lani, HI,
  USA: IEEE, 2017.

\bibitem{Guerreiro00}
B.~Guerreiro, C.~Silvestre, R.~Cunha, and A.~Pascoal, ``Trajectory tracking
  nonlinear model predictive control for autonomous surface craft,'' \emph{IEEE
  Transactions on Control Systems Technology}, vol.~22, no.~6, pp. 2160--2175,
  2014.

\bibitem{Svec00}
P.~Svec, A.~Thakur, B.~C. Shah, and S.~K. Gupta, ``Usv trajectory planning for
  time varying motion goals in an environmetn with obstacles,'' in
  \emph{Proceedings of the 2012 ASME International Design Engineering Technical
  Conference (IDETC)}.\hskip 1em plus 0.5em minus 0.4em\relax Chicago, USA:
  ASME, 2012.

\bibitem{Dolinskaya00}
M.~J. Hirsh, D.~Schroeder, A.~Maggiar, and I.~Dolinskaya, ``Multi-depot vessel
  routing problem in a direction dependent wavefield,'' \emph{Journal of
  Combinatorial Optimization}, vol.~28, no.~1, pp. 38--57, 2014.

\bibitem{Kim01}
H.~Kim, S.-H. Kim, M.~Jeon, J.~Kim, S.~Song, and K.-J. Paik, ``A study on path
  optimization method of an unmanned surface vehicle under environmental loads
  using genetic algorithm,'' \emph{Ocean Engineering}, vol. 142, pp. 616--624,
  2017.

\bibitem{NOAA00}
NOAA, ``How fast is the gulf stream?'' \emph{NOAA}, 2013.

\end{thebibliography}

\end{document}